\documentclass[twoside,leqno,twocolumn]{article}
\usepackage{ltexpprt}

\usepackage{graphicx}
\usepackage[latin1]{inputenc}
\usepackage{tikz}
\usetikzlibrary{shapes,arrows}
\usepackage{adjustbox}
\usepackage{gensymb} 

\usepackage{natbib}
\usepackage{eqnarray}
\usepackage{url}
\usepackage{amsmath,amssymb,amscd,graphicx}

\usepackage{subfigure}
\usepackage[colorlinks,citecolor=blue,urlcolor=blue]{hyperref}
\newcommand*\samethanks[1][\value{footnote}]{\footnotemark[#1]}

\begin{document}

\title{\Large Modeling Weather-induced Home Insurance Risks with Support Vector Machine Regression\footnote{Published in SIAM SDM 17 Workshop on Mining Big Data in Climate and Environment.} }

\author{Asim Kumer Dey\thanks{University of Texas at Dallas.}\\
\and
Vyacheslav Lyubchich\thanks{University of Maryland Center for Environmental Science.}\\
\and
Yulia R. Gel\samethanks[1]}

\date{}

\maketitle


\begin{abstract} \small\baselineskip=9pt
Insurance industry is one of the most vulnerable sectors to climate change. Assessment of future number of claims and incurred losses is critical for disaster preparedness and risk management. In this project, we study the effect of precipitation on a joint dynamics of weather-induced home insurance claims and losses. We discuss utility and limitations of such machine learning procedures as
Support Vector Machines and Artificial Neural Networks, in forecasting future claim dynamics and evaluating associated uncertainties. We illustrate our approach by application to attribution analysis and forecasting of weather-induced home insurance claims in a middle-sized city in the Canadian Prairies.
\end{abstract}

\section{Introduction.}

Changes in extreme weather and climate events, such as heat waves, droughts, heavy rainfall, and flood are the primary way that people experience climate change. Over the last 50 years, much of the United States has seen increases in prolonged periods of excessively high temperatures, droughts, and, in some regions, high precipitation and severe floods. Since 1950, extreme precipitation events have become more common in many regions in the United States \cite{NCEI}. Heavy precipitation is the primary cause of floods, when streams and rivers overflow their banks. Since 2008, the United States has seen six floods costing at least \$1 billion each, resulting in damaged property and infrastructure, agricultural losses, displaced families, and even loss of life \cite{c2es}, \cite{SOA}. Similarly to the USA, the number of flood disasters in Canada has noticeably increased in the past decades \citep{PSEPC}, \cite{Brooks}, \cite{c2es}.

Within the financial sector, the insurance industry faces the most substantial challenges from such climatic changes. From 2009 to 2014, in Canada, total insured losses from catastrophic events were close to or above \$1 billion each year---most
caused by
water damage. In fact, in 2013 Canadian insurers paid out a record-high \$3.2 billion to policy holders \cite{IBC}. The total rainfall related water damage incurred losses for the period of April--September in 1992--2002 sum up to about \$970 million for the four selected Canadian cities combined (London, Kitchener-Waterloo, Toronto, and Ottawa) \cite{Cheng:etal:2012}. The Intergovernmental Panel on Climate Change~(IPCC) has projected that the severity and frequency of extreme rainfalls (and, consequently, floods) will further increase \cite{IPCC}. Thus, the flood related losses will rise in the future, jeopardizing financial stability of the insurance industry and the whole society \cite{SOA}.

Various conventional statistical approaches, e.g., Generalized Linear Model (GLM), Autoregressive Integrated Moving Average (ARIMA) model, and Bayesian hierarchical model, have been employed for modeling the number of insurance claims (see the recent reviews by \cite{Haug:etal:2011}, \cite{Soliman:etal:2015}, \cite{Lyubchich:Gel:2017:insurance}, \cite{Scheel:etal:2013} and references therein).
More recently, different machine learning techniques, such as Neural Network (NN), Support Vector Machine (SVM), and Random Forest, have also been successfully utilized for assessing dynamics of claim frequencies \cite{Wu2011SupportVR}, \cite{CALDEIRA201562}.
Nevertheless, there exists a gap in understanding on how weather-induced losses, or dollar magnitudes of claims, depend on excessive
precipitation and how to \textit{simultaneously} model and forecast a joint dynamics of weather-related losses and frequencies.

We propose a new two-stage procedure to bridge the above gap based on Support Vector Regression (SVR) with Genetic Algorithm (GA), further referred to as GA-SVR. The main contributions of this paper are as follows:
\begin{itemize}
    \item We propose a new parsimonious two-stage approach for the \textit{joint} attribution analysis of weather-induced home insurance claims (frequencies) and losses (severities). We show that the new nonparametric attribution procedure delivers a competitive performance in characterizing spikes in the time series of claims and losses, while using a single exogenous regressor, i.e., precipitation.


    \item We use a computationally efficient data-driven GA approach to optimize hyper-parameters of SVR. We evaluate the new SVR-GA procedure for weather-induced risk attribution in respect to Artificial Neural Networks (ANN) and conventional SVR procedures. We find that SVR-GA leads to substantial gains in performance, that is, SVR-GA delivers at least 70\% lower root mean squared error for the number of claims and more than 10 times lower root mean squared error for aggregate losses than the competing approaches.

	\item We develop projected climate-induced numbers of claims and losses for 2021--2080 and investigate a relative change in expected claim dynamics vs. the insurance claims observed over a baseline period of 2002--2011. The forecasts are governed by the latest climate projections delivered by the Canadian Earth System Model (CanESM2) and downscaled by the Canadian Regional Climate Model (CanRCM4).



\end{itemize}

The remainder of this paper is organized as follows. In \S \ref{sec:Data} we describe insurance and weather data employed in our analysis. The GA-SVR technique is discussed in \S \ref{sec:SVR}. The results of modeling and forecasting are reported in \S \ref{sec:Modeling} and discussed in \S \ref{sec:Conclusion}.

\section{Data.}\label{sec:Data}
The data comprise of daily rainfall related home insurance claims and losses recorded in a middle-sized city in the Canadian Prairies (City~A) in the 10-year period of 2002--2011. The city name is suppressed due to data confidentiality. To remove the effects of changing risk exposure, the number of claims is normalized by the number of homes insured in City~A on each day from 2002 to 2011. To remove the effect of inflation, home insurance losses are converted to the prices of 2002 (CAD2002) using metropolitan area composite index of apartment building construction.
Future daily precipitation data for 2021--2080 are obtained from the CanESM2 global climate model downscaled to 0.22\degree\ horizontal grid resolution (approximately 25~km) by CanRCM4 model. Particularly, we use the data from the CanRCM4 grid cell which center is the closest to City~A. As the final pre-processing step, all daily data are aggregated to the weekly level.

Two scenarios of Representative Concentration Pathways (RCPs) available from CanRCM4 projections are RCP~4.5 and RCP~8.5. RCPs define possible climate futures 
depending on the amount and timing of greenhouse gases emissions in the future. RCP~4.5 assumes that the emissions will peak around 2040, then decline. Under the RCP~8.5 scenario, emissions continue to rise throughout the period 2000--2100~\cite{IPCC_RCP}. Projected precipitation amounts corresponding to these scenarios are used to predict future number of claims as well as aggregate loss. Table~\ref{tab:Overview} provides an overview of the data sets that are used in this study.

\setlength{\tabcolsep}{5pt}
\begin{table}
	\caption{Overview of the data sets}\label{tab:Overview}
	\small
	\begin{center}
		\begin{tabular}{l*{6}{c}r} \hline
			Period& Data type & Variables\\
			\hline
			Control & Observations  &precipitation, \\
            (2002--2011)                     &               &number of claims,\\
                                 &               &losses\\ [5pt]
			Scenario& Projections RCP 4.5& precipitation\\
            (2021--2080)                   & Projections RCP 8.5& precipitation\\
			\hline
		\end{tabular}
	\end{center}
\end{table}

\section{Support Vector Regression Approach.}\label{sec:SVR}

Support Vector Regression (SVR) has emerged as an alternative and highly effective tool for solving nonlinear regression and time series problems. Due to its wide applicability, SVR became popular in a variety of forecasting applications, including forecasts of warranty claims \cite{Wu2011SupportVR}, volatility of financial returns \cite{Chen2008}, water requirements \cite{JU2014523}, and tourism demand \cite{Chen2007}. Yuan~\cite{Yuan2012ParametersOU} uses SVR with genetic algorithm to forecast sales volume.

\begin{figure}
    \centering
    \includegraphics[width=.99\linewidth]{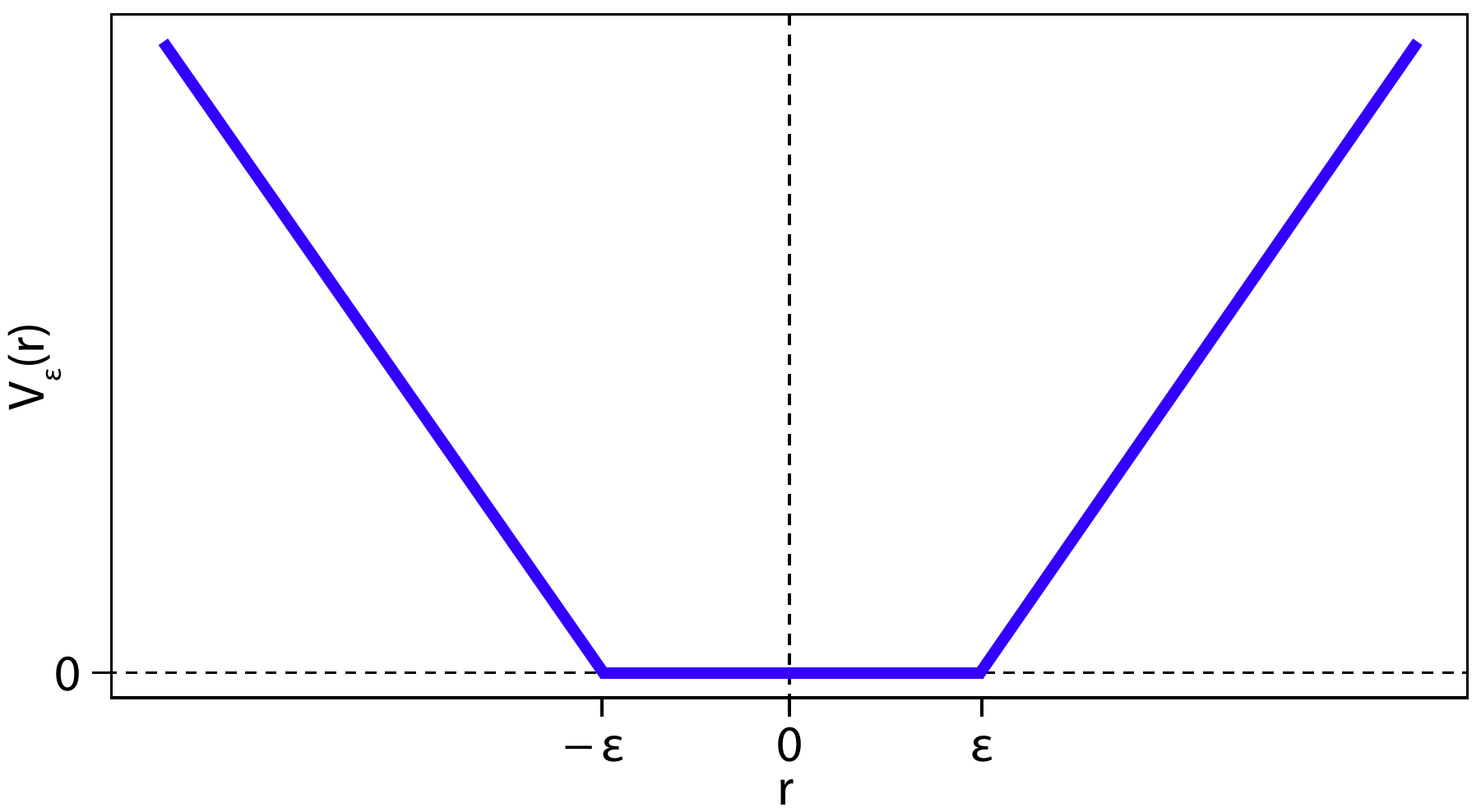}
    \caption{SVR $\varepsilon$-insensitive error function.}\label{fig:ErrorFun}
\end{figure}

SVR, a version of Support Vector Machine (SVM)~\cite{NIPS1996_4f284803}, is based on the statistical learning theory \cite{vapnik74theory}. In contrast to Least Squares Regression, where  in-sample residual sum of squares is minimized, SVR attempts to minimize the generalized error bound. This bound is the combination of the training error and a regularization term that controls the complexity of the hypothesis space \cite{Basak2007}. Another important feature of SVR is that the method only depends on a particular 
subset of the training data, because only residuals that are larger than an absolute value of some constant ($\varepsilon$) contribute to the loss function. For this reason, the SVR loss function is called $\varepsilon$-insensitive loss function (Figure~\ref{fig:ErrorFun}).

Let
$D = \{(x_i,y_i)\}$ be a learning sample set, where $x_i \in \mathcal R^m$ represents input values; $y_i \in \mathcal R$ are corresponding output values for $i= 1, 2, \ldots, N$; $N$ is sample size, and $m$ is dimension of the input dataset. The aim is to identify a regression function, $y=f(x)$, that accurately predicts the outputs $\{y_i\}$ corresponding to a new data set $D$. 
In the case of a linear SVR, 
$f(\cdot)$ has the form
\begin{equation}\label{eq:SVM}
f(x)= 	\langle\beta, x \rangle + \beta_0,  
\end{equation}
where $\beta \in  \mathcal R^m$, $\beta_0 \in \mathcal R$, and $\langle \cdot, \cdot \rangle$ denotes inner product in $R^m$. To estimate  $\beta$ and $\beta_0$, we consider minimization of the following risk function:
\begin{equation}\label{eq:RiskFun}
H(\beta, \beta_0)=\sum\limits_{i=1}^n V_{\varepsilon}\left(y_i-f(x_i)\right)+ {\lambda \over 2}{\|\beta\|}^2,
\end{equation}
where $V_{\varepsilon}(r)$ is $\varepsilon$-insensitive error function 
(Figure~\ref{fig:ErrorFun}),
$\lambda>0$ is regularization constant, and
${\|\cdot\|}^2$ denotes Euclidean norm \cite{Hastie:etal:2009}, \cite{GHORBANI2016301}.

The risk function, $H(\beta, \beta_0)$, involves summation of the empirical risk (training errors) and regularization term ${{\|\beta\|}^2 / 2}$. The regularization parameter, $\lambda$, determines the trade-off between minimizing empirical risk and minimizing the regularization term.
Minimization of 
function (\ref{eq:RiskFun}) leads to a quadratic programming problem, which can be converted to the dual Lagrangian problem \cite{Hastie:etal:2009}, \cite{Chen2007}. Its solution function 
has the form~\cite{Basak2007}:
\begin{equation}
\hat{f}(x) = \sum\limits_{i=1}^n (\hat{\alpha}_{i}^\ast -\hat{\alpha}_{i}){\langle x, x_i\rangle} + \beta_0,
\end{equation}
 where $\hat{\alpha}_{i}^\ast$  and  $\hat{\alpha}_{i}$ are positive.

The linear SVR (\ref{eq:SVM}) can be extended to nonlinear regression through a mapping function $\phi(x)$. We can map every input data  point into a high-dimensional feature space through the nonlinear mapping function, $\phi(x)$, and then apply the standard SVR algorithm: 
\begin{equation}\label{eq:NLSVM}
\hat{f}(x) = \sum\limits_{i=1}^n (\hat{\alpha}_{i}^\ast -\hat{\alpha}_{i}){\langle \phi(x), \phi(x_i)\rangle} + \beta_0.
\end{equation}

The inner product in (\ref{eq:NLSVM}) can be replaced by a kernel function $k(x_i, x)$. The kernel function can handle any dimension feature space \cite{Tay2001}. Different types of kernel functions are used in SVR. In this study, we use the Radial Basis Function (RBF) kernel (Gaussian kernel):
\begin{equation}\label{eq:kernel}
k(x_i,x)=\exp \left({-{{{\|x-x_i\|}^2} \over {2{\sigma^2}}}}\right),
\end{equation}
where $\sigma^2$ is the bandwidth. 

However, results of SVR modeling highly depend on a number of user-defined parameters (hyper-parameters), which should be carefully selected. Such parameters include:
\begin{itemize}
  \item Regularization parameter ($\lambda$),
  \item Tube size of $\varepsilon$-insensitive loss function ($\varepsilon$),
  \item Bandwidth of the kernel function ($\sigma^2$).
\end{itemize}

\begin{figure}
	\tikzstyle{decision} = [diamond, draw,  text width=4.5em, text badly centered, node distance=3cm, inner sep=0pt]
	\tikzstyle{block} = [rectangle, draw,text width=15em, text centered, rounded corners, minimum height=4em]
	\tikzstyle{line} = [draw, -latex']
	\tikzstyle{cloud} = [draw, ellipse, node distance=3cm, minimum height=2em]
	
	\begin{adjustbox}{max totalsize={.5\textwidth}{.4\textheight},center}
		
		\begin{tikzpicture}[node distance = 2cm, auto]
		\node [block] (init) {Define:\\
			- Parameters ($C$, $\sigma$, $\varepsilon$)\\
			- Population size\\
			- Fitness function\\
			- Stopping criteria };
		
		\node [block, below of=init, node distance=2.5cm] (identify) {Generate initial random population};
		\node [block, below of=identify, node distance=2.5cm] (Train) {Train SVR model and calculate fitness};
		
		\node [decision, below of=Train, node distance=2.3cm] (decide) { $i=n$?};
		\node [block, right of=decide , node distance=7cm] (update) {Create new population by:\\
			- Reproduction\\ - Crossover\\ - Mutation};
		\node [block, below of=decide, node distance=2.5cm] (stop) {Select optimal ($C$, $\sigma$, $\varepsilon$)};
		\node [block, below of=stop, node distance=2.1cm] (train1) {Train SVR model using obtained hyper-parameters };
		\path [line] (init) -- (identify);
		\path [line] (identify) --  node {$i=1$}(Train);
		\path [line] (Train) -- (decide);
		\path [line] (decide) --node{No} node[anchor=north] {$i=i+1$}(update);
		\draw [line] (update) |- (Train);

		\path [line] (decide) -- node {Yes}(stop);
		\path [line] (stop) -- (train1);
		\end{tikzpicture}
	
	\end{adjustbox}
	\caption{Flow-chart of the GA-SVR.}\label{fig:GASVR}
\end{figure}
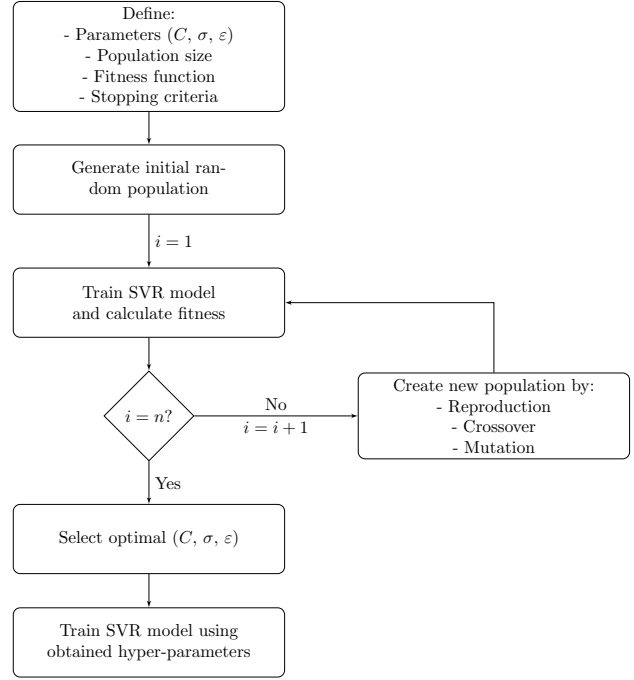

Inappropriate choice of SVR parameters may lead to over-fitting or under-fitting \cite{Lin2005SupportVR}. An unconnected, separate optimization is not sufficient for finding the optima because these variables influence each other. Different methods can be employed for selecting hyper-parameters (see \cite{Chen2007} for an overview). In this study, Genetic Algorithm (GA) \cite{goldberg89}, \cite{Deepa2008}  is applied to simultaneously optimize all SVR parameters. GA is a global search procedure: it checks all solution space from many points for the best combination of free parameters. A general procedure of GA for selecting the optimal hyper-parameters is shown in Figure~\ref{fig:GASVR}. Genetic algorithm uses selection, crossover, and mutation operations to generate the offspring of the existing population \cite{Scrucca2013}, \cite{Yuan2012ParametersOU}. When convergence criteria are met, the algorithm (evolution) stops and delivers the hyper-parameter values corresponding to the best fitness function value.

\section{Modeling Precipitation Related Risk.}\label{sec:Modeling}

In our study, the number of claims and aggregate loss are modeled in a two-stage procedure, using Genetic Algorithm Support Vector Regression (GA-SVR).
The mapping function for GA-SVR for the number of claims is
\begin{equation}\label{eq:GASVRclaims}
N_t=f{(R_t, R_{t-1}, \max R_t)}
\end{equation}
and for the aggregate loss
\begin{equation}\label{eq:GASVRloss}
L_t=f{(R_t, R_{t-1}, \max R_t, N_t),}
\end{equation}
where $N_t$ is number of claims at week $t$, $L_t$ is aggregate loss at week $t$,  $R_t$ is total precipitation at week $t$, and $\max R_t$ is maximum daily total precipitation at week $t$.
GA parameter settings used in this study are given in Table~\ref{tab:GAparameter}.

\setlength{\tabcolsep}{5pt}
\begin{table}
   \caption{GA parameter settings}\label{tab:GAparameter}
   \small
	\begin{center}
	\begin{tabular}{llccc} \hline
\multicolumn{2}{l}{Parameter} & Value \\
\hline
		\multicolumn{2}{l}{Number of generations} & \\
\multicolumn{2}{l}{(stopping criterion):}&100\\
\multicolumn{2}{l}{Population size:} & 50 \\
\multicolumn{2}{l}{Fitness function:} & RMSE\\
Search & $C$\hspace{2em} & ($10^{-3}$, $10^3$)\\
 domain \cite{HsuLibsvmTutorial2003}: &${\sigma}^2$ & ($10^{-3}$, $2^4$) \\
 &$\epsilon$ &(${10}^{-2}$, $2^3$)\\ \hline
	\end{tabular}
\end{center}
\end{table}

\subsection{Prediction.}

We fit models (\ref{eq:GASVRclaims}) and (\ref{eq:GASVRloss}) using the data from the control period (see Table~\ref{tab:Overview}) and obtain expected values of weekly number of claims ($\hat{N}_t$) and weekly aggregate loss ($\hat{L}_t$) using the RCP~4.5 and RCP~8.5 data for the scenario period.
As the aggregate loss model relies on the number of claims $N_t$, we start from forecasting future number of claims. Due to the large uncertainties of climate projections at such high---weekly---temporal resolution (see \cite{Lyubchich:Gel:2017:insurance} on the potential sources of uncertainties), we consider aggregated trends in the future, rather than individual weekly number of claims or aggregate loss. We summarize the trends as the relative change (measured in percents) from the control period to the projection (scenario) period. For balanced comparisons of scenario with control, we split the 60-year scenario period (2021--2080) into six 10-year sub-periods to make the control period (2002--2011) and scenario periods (2021--2030, 2031--2040, $\ldots$, 2071--2080) of equal length. Then, we calculate the percentage change in the number of claims or aggregate loss for each of the six scenario sub-periods: 
\begin{equation}\label{eq:Delta}
\Delta=\bigg({{\sum_{t\in scn} \hat{N}_{t}} \over {\sum_{t\in ctr} {N_{t}}}} -1 \bigg) 100 \%,
\end{equation}
where $scn$ and $ctr$ refers to the scenario sub-period and control period, respectively. Since both periods are of 10 years, the numerator and denominator are of comparable size, so no further normalization is necessary.

\begin{figure}[h]
\centering
\subfigure[Number of claims]{
	    \includegraphics[width=0.48\textwidth, viewport= 0 15 550 378, clip]{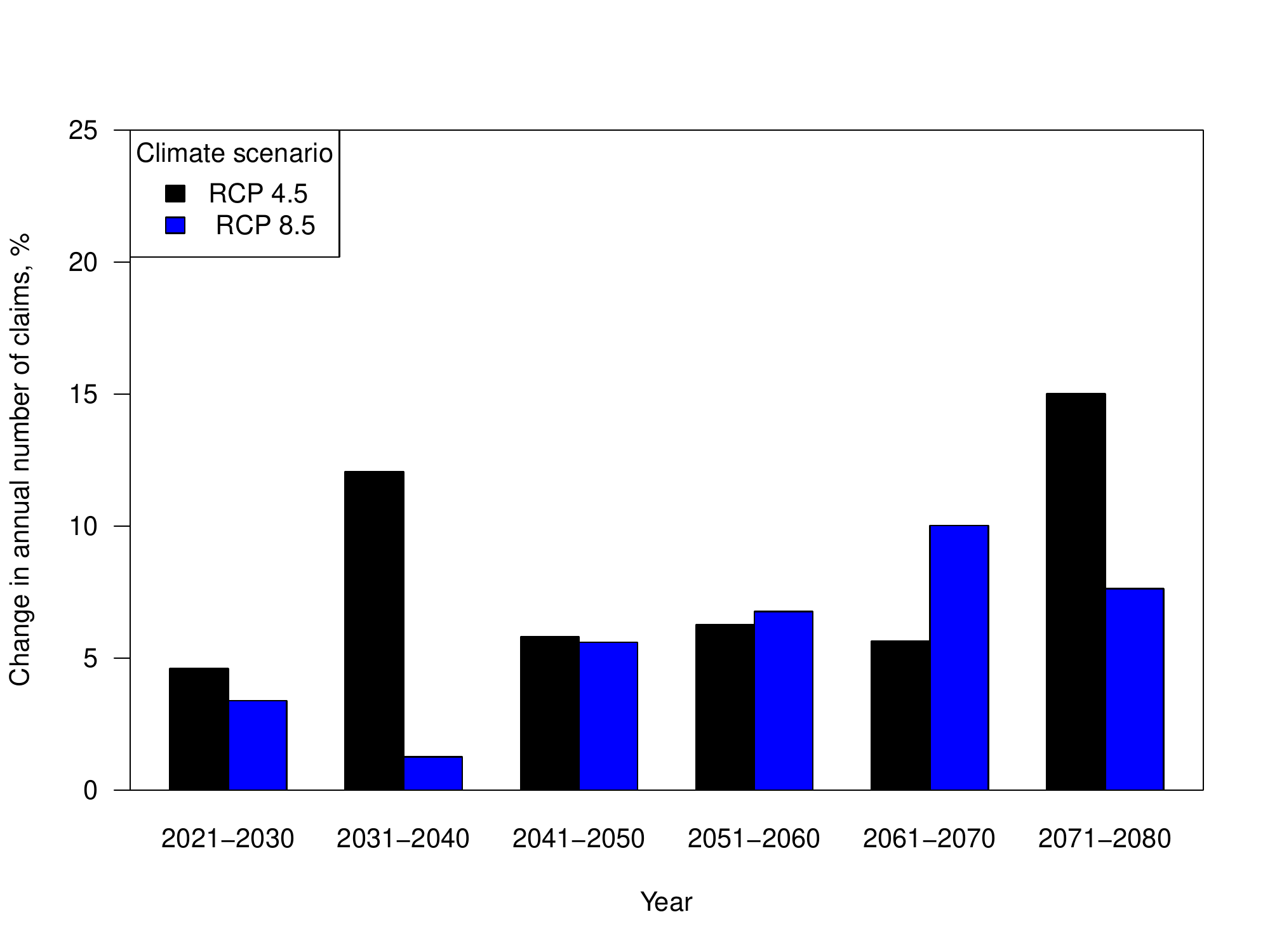}
 	    \label{fig:DeltaClaims}
	}\\ 
\subfigure[Aggregate loss]{
	    \includegraphics[width=0.48\textwidth, viewport= 0 15 550 378, clip]{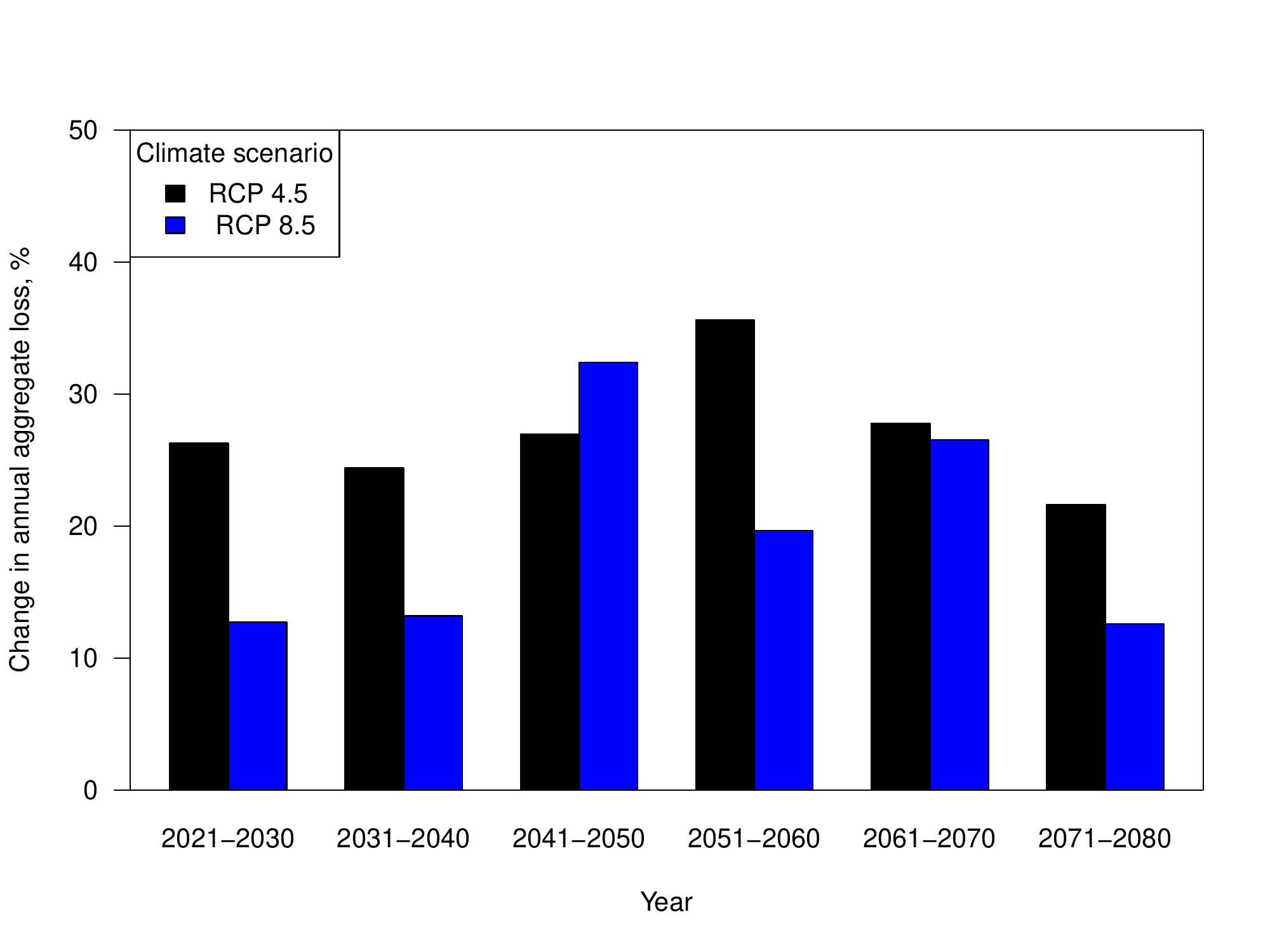}
 	    \label{fig:DeltaLosses}
	}	
	\caption{Projected percentage changes in home insurance claims and losses in City A, relative to the control period of 2002--2011.}
    \label{fig:Delta}
\end{figure}

We now assess the
expected claim and aggregate loss dynamics corresponding to the two climate change scenarios for the area of City A.
Figure~\ref{fig:DeltaClaims} shows that the annual number of home insurance claims will increase noticeably in the scenario period, relative to the baseline of 2002--2011. The projected number of claims for the climate scenario RCP~4.5 
is from 4.5\% to 15.0\% higher than in 2002--2011. Under the scenario RCP~8.5, number of claims is expected to be from 1.3\% to 10.0\% higher than in the baseline period.

The magnitude of projected increase in annual aggregate loss for the same city is substantially higher, which is from 21.6\% to 35.6\% for the scenario RCP~4.5, and from 12.7\% to 32.4\% for the scenario RCP~8.5 (Figure~\ref{fig:DeltaLosses}). Such mismatch in the relative change of the number of claims and losses (i.e., total losses grow faster than the number of claims)  might be attributed, for example, to an increased future number of high impact atmospheric events such as major floods, and/or major weather-related breakdowns of outdated civil infrastructure.
In general, we can conclude
that the average climate-induced loss is expected to rise regardlessly of the Representative Concentration Pathways (RCPs) scenario.


\subsection{Goodness of Fit.}

In this section, we validate the employed GA-SVR models by comparing their goodness of fit with the following highly competitive approaches:
\begin{itemize}
  \item SVR without automatic selection of the optimal hyper-parameters,
  \item Artificial Neural Network (ANN).
\end{itemize}

\begin{figure}	
\begin{adjustbox}{max totalsize={.6\textwidth}{.5\textheight},center}
\tikzset{
	every neuron/.style={
		circle,
		draw=black,	fill = white,
		minimum size=.45cm
	},
	neuron missing/.style={
		draw=none,
		scale=6,
		text height=0.233cm,
		execute at begin node=\color{black}$\vdots$
	}
}

	\begin{tikzpicture}
	
	\foreach \m/\l [count=\y] in {1,2,3,4}
	\node [every neuron/.try, neuron \m/.try] (input-\m) at (0,2.5-\y) {};
	
	\foreach \m [count=\y] in {1,2}
	\node [every neuron/.try, neuron \m/.try ] (hidden-\m) at (2,2-\y*1.25) {};
	
	\foreach \m [count=\y] in {2}
	\node [every neuron/.try, neuron \m/.try ] (output-\m) at (4,1-\y) {};
	
	\foreach \l [count=\i] in {1,2,3,4}
	\draw [<-] (input-\i) -- ++(-1,0)
	node [above, midway] {$X_\l$};
	
	\foreach \l [count=\i] in {1,2}
	\node [above] at (hidden-\i.north) {$Z_\l$};
	
    \foreach \i in {2}
	\draw [->] (output-\i) -- ++(1,0)
	node [above, midway] {$Y$};
	
	\foreach \i in {1,...,4}
	\foreach \j in {1,2}
	\draw [->] (input-\i) -- (hidden-\j);

	\foreach \i in {1,2}
	\foreach \j in {2}
	\draw [->] (hidden-\i) -- (output-\j);

	\foreach \l [count=\x from 0] in {Input, Hidden, Output}
	\node [align=center, above] at (\x*2,2) {\l \\ layer};

	\end{tikzpicture}
	\end{adjustbox}
	
	\caption{Neural network for four inputs and one hidden layer with two units.}
	\label{f:figNN}
\end{figure}
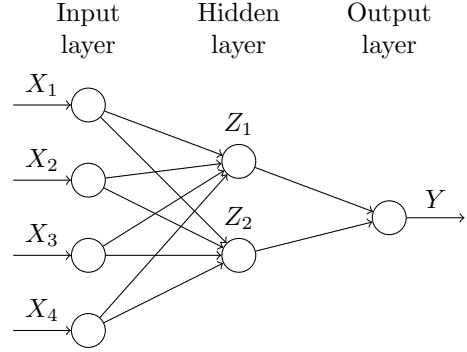

A neural network is a two-stage regression model, typically represented by a network diagram. The central idea is to extract linear combinations of the inputs as derived features, and then model the target as a nonlinear function of these features. Let $X_1$, $X_2$, $X_3$, and $X_4$ be the four input variables (like in the loss model (\ref{eq:GASVRloss})), $Z_1$ and $Z_2$ be the two hidden units, and $Y$ be the output. A network diagram for this model is shown in Figure~\ref{f:figNN}, where
\begin{eqnarray} \nonumber
Z_j&=&f\left(\sum\limits_{i=1}^4 {\alpha_{ij}}{X_i}+ \alpha_{oj}\right), \qquad f(v)={1 \over {1+e^{-v}}}, \\
Y&=&g\left(\sum\limits_{j=1}^2{\beta_{j}}{Z_j}+ \beta_{o}\right) ,  \qquad \quad \! g(v)={v}.\nonumber
\end{eqnarray}

The activation functions $f(\cdot)$ and $g(\cdot)$ used in this study are sigmoid and identity activation functions, respectively. The unknown parameters (weights) $\alpha_{ij}$, $\alpha_{oj}$, $\beta_{j}$, and $\beta_{o}$ are estimated by back-propagation method, with gradient descent updated iteratively \cite{Bishop2006}, \cite{Hastie:etal:2009}.

\begin{figure}[t!]
 	\centering
    \subfigure[Number of claims]{
	    \includegraphics[width=0.48\textwidth, viewport= 0 15 550 378, clip]{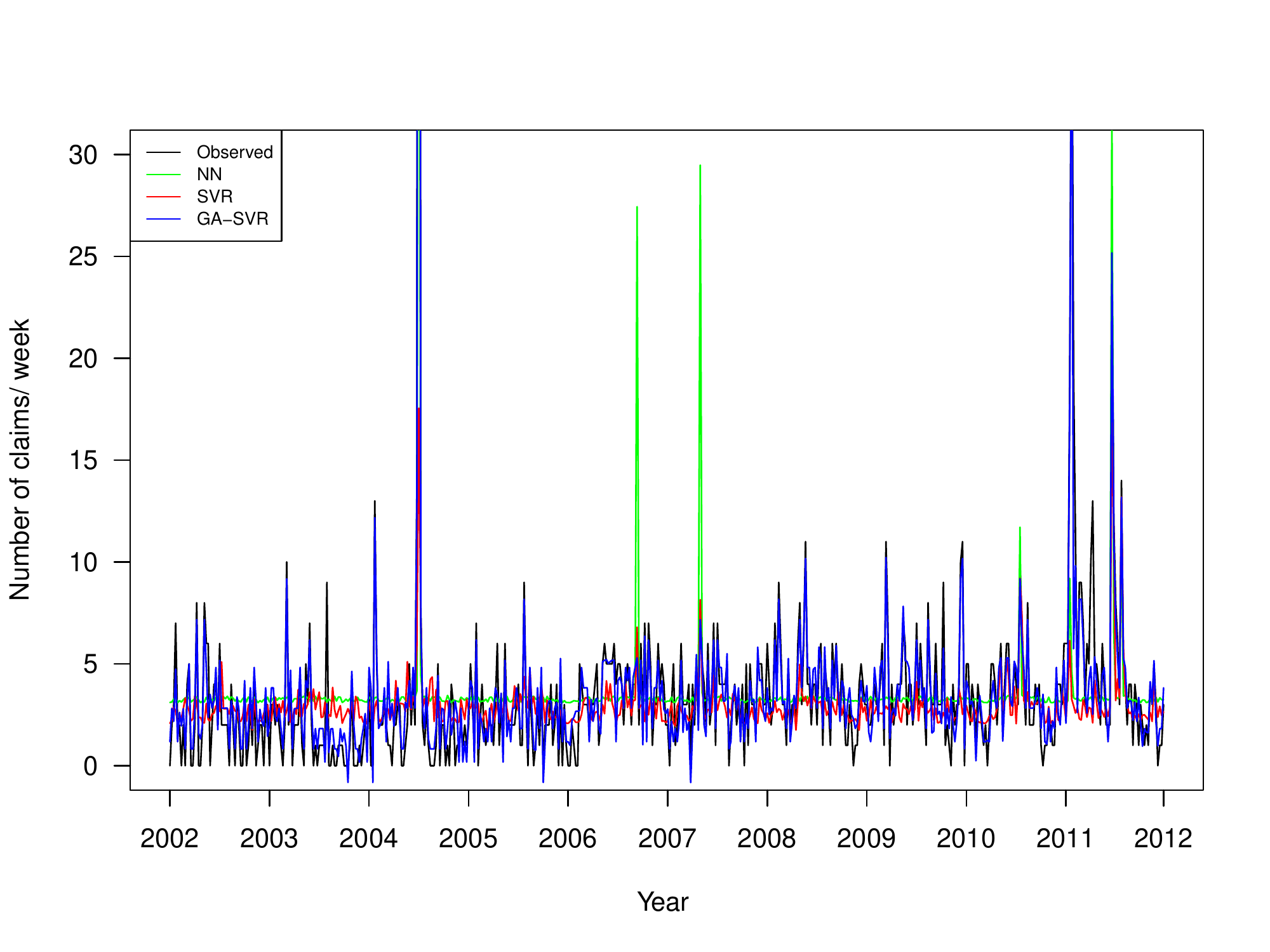}
 	    \label{fig:MultimodelsClaimsA}
	}
 	    \subfigure[Total loss]{
	    \includegraphics[width=0.48\textwidth, viewport= 0 15 550 378, clip]{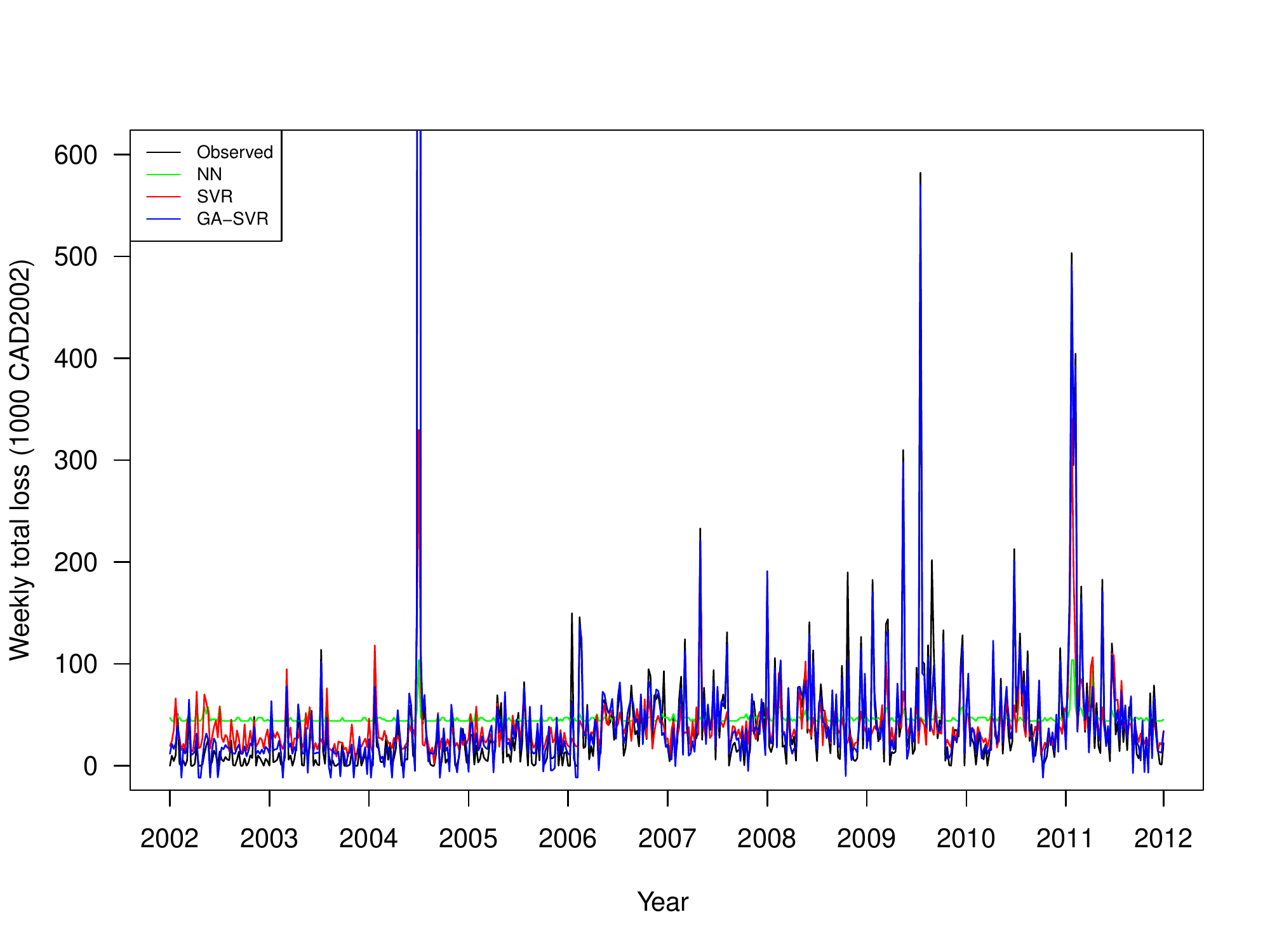}
 	    \label{fig:MultimodelsLossesA}
	}
 	\caption{Observed and fitted with three different models weekly time series of home insurance 
             claims and 
             losses in City A.
 }
    \label{fig:Multimodels}
 \end{figure}

Table~\ref{tab:RMSEclaims} shows that root mean squared error (RMSE) of GA-SVR approach for modeling the number of claims, $N_t$, is much smaller than of ANN or SVR.
A substantial advantage of the GA-SVR is that the procedure better captures the variability of observed data, particularly, the sudden high spikes in the number of claims, which are the paramount concern for insurance companies (Figure~\ref{fig:MultimodelsClaimsA}).

\begin{table}[t!]
	\caption{Training RMSE of the models for the number of claims ($N_t$) and aggregate loss ($L_t$)}\label{tab:RMSEclaims}\label{tab:RMSElosses}
	\begin{center}
		\begin{tabular}{lrr} \hline
			Model & $N_t$ & $L_t$\\
			\hline
			ANN & 5.25 &  212858.4\\
			SVR & 7.32 & 200660.1 \\
			GA-SVR & 1.37& 13955.6 \\
			\hline
		\end{tabular}
	\end{center}
\end{table}


SVR-GA delivers at least 70\% lower root mean squared error (RMSE) for the number of claims than the competing approaches, ANN and SVR (see Table~\ref{tab:RMSElosses}).
Furthermore, GA-SVR model also yields the most accurate fit to the observed aggregate losses, $L_t$, i.e., RMSE of GA-SVR is up to 15 times lower than of RMSE of ANN and SVR.
Finally, GA-SVR also captures variability of the observed total losses (see Figure~\ref{fig:MultimodelsLossesA}).



\section{Conclusion.}\label{sec:Conclusion}

The purpose of this study is to model and forecast dynamics of weather-induced home insurance claims and total losses in respect to changes in precipitation. We employ support vector regression with genetic algorithm (GA-SVR)---a machine learning procedure that allows for flexible data-driven characterization of claim dynamics as a function of atmospheric information. We illustrate our approach in application to insurance claims in a middle-sized city in the Canadian Prairies.

Our model is based on a two-stage approach, that is, we first model a number of claims as a function of precipitation and then assess dynamics of total losses as a function of claim frequency and precipitation.
We evaluate sensitivity of projected claims in respect to different greenhouse effect scenarios (RCP~4.5 and RCP~8.5). The projected number and magnitude of claims are compared to the baseline (control) period of 2002--2011. Our results
indicate that in the future we can expect that both number of claims and insurance losses will increase in City~A, but the rate of increase depends on the climate change scenarios.

It is important to emphasize that the obtained projections of future number of claims and losses account only for changes in future precipitation rather than other meteorological and non-meteorological factors, e.g., seasonal component, socio-economic changes, location and value of assets, that could also influence the dynamics of insurance claims. In the future, we plan to incorporate such information into assessment of future claim dynamics.

Predictions of future insurance claims and losses embody various sources of uncertainty. First, there is the uncertainty inherent in the climate model projections. Second, there is uncertainty about the specified models for insurance claims and losses. The third source of uncertainty is from estimates of the model parameters. In the future, we plan to quantify different sources of uncertainty. We also plan to expand the spatial domain of our analysis by evaluating the dynamics of weather related home insurance in other cities.

\bibliographystyle{abbrv}
\bibliography{RefInsurance}

\end{document}